\documentclass[a4paper,10pt]{extarticle}
\usepackage[utf8]{inputenc}
\usepackage{amsmath}
\usepackage{amssymb}
\usepackage{graphicx}
\usepackage{algpseudocode}
\usepackage[noend,ruled,vlined]{algorithm2e}
\usepackage[margin=1in]{geometry}
\usepackage{graphics}
\usepackage[para]{footmisc}

\title{Obfuscation of Discrete data}
\author{Saswata Naha\footnote{} \\Sayantan Roy\footnote{}\\ Arkaprava Sanki\footnote{}\\Diptanil Santra\footnote{Indian Statistical Institute Kolkata}}
\date{June 2021}

\begin{document}

\maketitle

\begin{abstract}
    \noindent Data obfuscation deals with the problem of masking a data-set in such a way that the utility of the data is maximized while minimizing the risk of the disclosure of sensitive information. To protect data we address some ways that may as well retain its statistical uses to some extent. One such way is to mask a data with additive noise and revert to certain desired parameters of the original distribution from the knowledge of the noise distribution and masked data. In this project, we discuss the estimation of any desired quantile and range of a quantitative data set masked with additive noise.
\end{abstract}

\vspace{1cm}

\section*{Introduction} 
 Privacy protection and data security have recently received a substantial amount of attention due to the increasing need to protect various sensitive information like economic data and medical data. In case of continuous data, a very famous example is publication of marks of students. An Institute would like to publish the performance of its students without disclosing the actual marks. Whereas, for discrete case, data on economic class study can be used, where it is of national interest to know about the economic status of the country, not disclosing the sensitive information of income of any individual. We are interested in the obfuscation of discrete case only and the desired estimations from such a data.

\vspace{5mm}

\section*{Basic Problem}
Our main goal through this project is to construct a way of obfuscation of discrete data using some additive noise in such a way that the probability of estimating the actual data given the obfuscated data is quite low. However, given the nature of the noise added to the original data, we can estimate certain population statistics with low error.
\;

Suppose, $X$ be the data vector containing the individual values of the population. We want to publish the distribution of $X$, $p_i=\mathbf{P}(X=x_i)\:\forall\:x_i\in$ range of $X$. For additive model, we add noise vector $Y$ with $q_j=\mathbf{P}(Y=y_j)\:\forall\:y_j\in$ range of $Y$ depending on the range of $X$ and get the masked data $Z=X+Y$. Our aim is to adjust $Z$ accordingly to publish the obfuscated data $Z'$ such that $\mathbf{P}(X\mid Z')\approx0$.\;

As for the next part of the project we want to estimate certain statistical properties of the true data viz. \textbf{Median} or other quantiles and \textbf{Range}. For this, we would like to estimate the actual probabilities $p_0,p_1,\cdots$.\;

Note that, $\displaystyle\mathbf{P}(Z=z_k)=\displaystyle\sum_{i=0}^k\mathbf{P}(X=x_i)\mathbf{P}(Y=z_k-x_i)=\sum_{i=0}^kp_iq_j$. Estimating $\mathbf{P}(Z=z_k)$ from the obfuscated data, we can try to solve for $p_i$'s from the given data of $q_j$'s.\\

\section*{Dataset}
 We have generated some poisson datasets and added some noise to them. Then truncated that in such a way that the probability of estimating the actual data from the obfuscated data is quite low.
\;

At first, 10 lakhs data from poisson with parameter generated from exponential (2.5).\\

\begin{center}
    \includegraphics[scale=0.6]{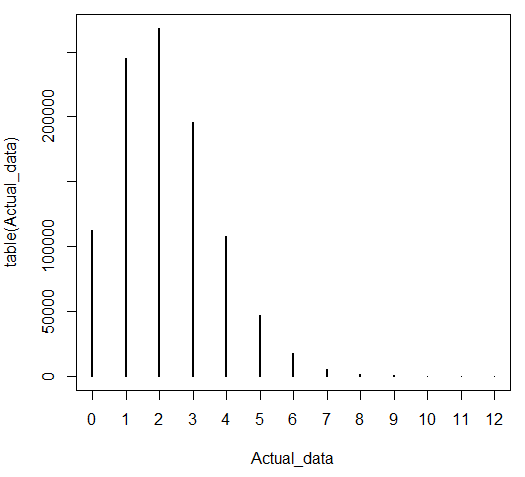} 
\end{center}

As we have mentioned before that economic data are more sensitive in nature and instead of publishing the actual data we want to obfuscate it before publishing it. This kind of data usually take positive values and they are positively skewed. We have generated the data in such a way that it can be compared with an economic data. Here in the data set, the variable in the X-axis denotes different economic class. The whole population is divided into 12 economic classes where $0^{th}$ class denote the lowest income group and $12^{th}$ class denote the highest income group. The variable in Y-axis denotes the frequency of each class. Then we obfuscate it in the following way:

We add an additive noise from unif(0,10).\\

\begin{center}
    \includegraphics[scale=0.6]{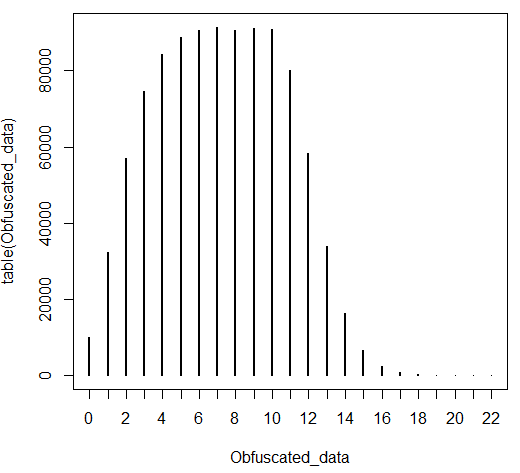}
\end{center}

Then we collect the household size data set from the website of ``Census of India'' and have used the data for the whole Indian region.\;

\pagebreak

The data is shown in the diagram below -

\begin{center}
    \includegraphics[scale=0.6]{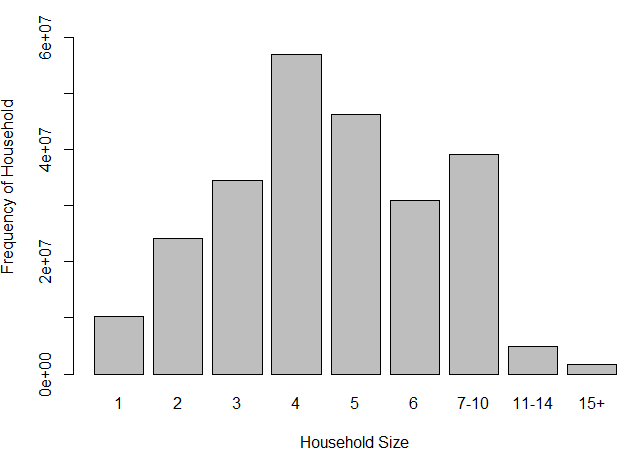}
\end{center}

Here X-axis denotes the household size and Y-axis denotes the  frequency corresponding to each household size.\;

Then we similarly add an additive noise from unif(1,10).\; 

\begin{center}
    \includegraphics[scale=0.6]{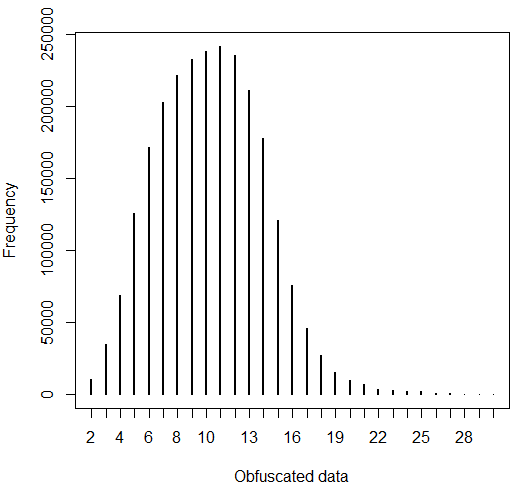}
\end{center}

\section*{Goodness of Obfuscation}
In this section, we want to show that using this obfuscation procedure, probability of getting back the actual data from the published obfuscated data is quite low.\;

Let X,Y,Z be the values of the actual data, added noise and values of obfuscated data respectively. Now, we have to calculate $P(X|Z)$. We know,

\[\mathbf{P}(X=x|Z=z) =\frac{\mathbf{P}(X=x,Z=z)}{\mathbf{P}(Z=z)}\]. \;

To calculate the probabilities in the numerator and denominator separately, first we construct a matrix with the values of actual data (X) in the rows and the corresponding error values (Y) in the columns. So the $(i,j)^{th}$ element of the matrix represents number of individuals in class \textbf{i} has taken error values \textbf{j}. So the size of the matrix is $12\times 10$. For our calculation, as the values of X have been fixed as x, the row wise sum of the matrix is fixed for all rows. Also note that, the sum of the off-diagonals of the matrix represents the values of Z. Fixing these values according to our data, we get the total number of possible cases, from which we can calculate the exact probability.\;

\;

Next, for the probability in the denominator we take all possible cases for X and Y such that the data in our case is valid in real situation. The computations are very cumbersome$^*$, however, the actual conditional probability turns out to be of the order $\mathcal{O}(10^{-10^6})$. So we can conclude that, since this probability in itself is very small, if we further mask the obfuscated data, the probability of getting the true data conditioned on that will be even smaller. So this method of obfuscation can be said to be satisfactory.\;

[* Cumbersome in the sense that, we have made an algorithm to count the number of matrices which satisfy the required conditions. In this way we got the proportion of matrices satisfying the conditions which will be close to the required probability.]

\section*{Different approaches to estimate quantiles}

Quantiles are robust statistics and hence is not much affected by the presence of outlying data-points. To do the estimation of quantiles we first estimate the c.d.f. of X from the obfuscated data using different approaches. \; 
 
\subsection*{First Approach}
First we try the most common method to estimate quantiles that is, using \textbf{least square} method from the complete obfuscated data. Moreover, we know that the least square method gives an unbiased estimate of parameter. So we shall use this in the following way:\;

We know that for a least square model, 
\begin{center}
    $\Tilde{Y}=X\Tilde{\beta}+\Tilde{\epsilon},$
    where, $\Tilde{\epsilon}\sim \mathcal{N}(\Tilde{0},\sigma^2I)$,
\end{center}
\;

We get the solution as $\hat{\beta}=(X'X)^{-1}X'\Tilde{Y}$.\;

However, as the components of $\Tilde{\beta}$ are probabilities and add up to 1, implementing the constraint of $\Tilde{\beta}'\Tilde{1}=1$, we get the solution as 
\begin{center}
    $\hat{\beta}=(X'X)^{-1}(X'\Tilde{Y}-\frac{\Tilde{1}'(X'X)^{-1}X'\Tilde{Y}-1}{\Tilde{1}'(X'X)^{-1}\Tilde{1}}\Tilde{1})$.
\end{center}

The result for our data set is like the following -\\
\begin{center}
    
    \includegraphics[scale=0.65]{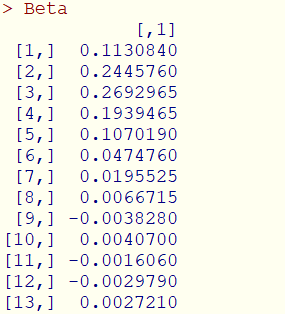}
\end{center}

[Necessary R codes are given in Appendix section.]\;

\;

But in this approach,as can be seen in the estimated solution, some of the components come out to be negative and hence cannot be probabilities. So, we need a method to further implement the non-negativity criteria of the solution or need to change the published data in some suitable way.\;

First we try to modify the solution method keeping the whole obfuscated data. We suspect that the inverse function in numerical methods can affect the solution to a certain effect. So, we then try using QR decomposition to check if there is any improvement. But some of the probabilities are found to be still negative. So, we conclude that, we need to rectify the approach itself.\;

\subsection*{Second Approach}

Since the first approach failed, we shall try MLE though it is not unbiased. But we know that MLE maximizes the likelihood (i.e, probability of the value of the parameter is the maximum according to the sample) and consistent. Also MLE belongs to the parameter space. Therefore, our next attempt is using \textbf{MLE} with truncation. Suppose, $\Tilde{p}$ and $\Tilde{r}$ be the probability distributions of the actual data and the obfuscated data respectively.\\
Maximizing the likelihood w.r.t $\Tilde{r}$ we get the MLE estimates as $\displaystyle r_i=\frac{n_i}{\Sigma n_j}\:\forall\:i$.\;

\noindent So, solving w.r.t $\Tilde{p}$ we get-\;

\begin{center}
$\displaystyle p_0=\frac{n_0}{\sum n_i}*11$, $\displaystyle p_j=\frac{n_j-n_{j-1}}{\sum n_i}*11$ for j = 1,2,...,11 and $\displaystyle p_{12}=1-\sum p_j$.
\end{center}

Similarly as the above, we can also estimate $\Tilde{p}$ as -\;
\begin{center}
$\displaystyle p_{12}=\frac{n_{22}}{\sum n_i}*11$, $\displaystyle p_j=\frac{n_{j+11}-n_{j+10}}{\sum n_i}*11$ for j = 1,2,...,11 and $\displaystyle p_0=1-\sum p_j$
\end{center}

Using R, both of the estimates are calculated, the results are shown below -

\begin{center}
    \includegraphics[scale=0.65]{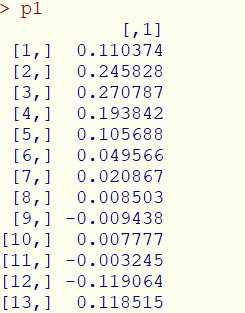}
    \includegraphics[scale=0.65]{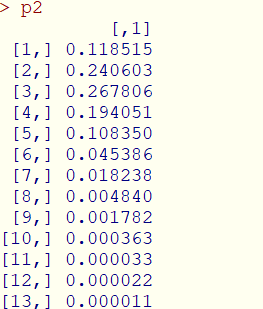}
\end{center}

[Necessary R codes are given in Appendix section.]\;

\;

However, in this case as well, some of the estimates turn negative and one way to deal in such situation is combine the above to estimates to get a better estimate.\;

A general technique in this combination is depending on data. Take that part of first estimate where the height of the columns of obfuscated data are increasing and that part of second estimate where the height of columns are decreasing.\;

For our data set, the combined estimate is as following -
\begin{center}
    \includegraphics[scale=0.65]{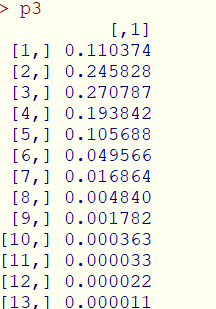}
\end{center}

[Necessary R codes are given in Appendix section.]\;

\;

 Now, there is two issues regarding this method - First of all, this method is on non-truncated obfuscated data, where there is a data leak in the extreme values. Secondly, the technique of the combination as mentioned above, failed to work if there are ups and downs in frequency all over the range.\;
 
 To avoid the first issue, we shall work on the truncated obfuscated data and for the second issue, we have to somehow extend our MLE approach to Constrained MLE.\;
 
 \subsection*{Third Approach}
 For the constrained MLE we need to include the constraints $p_i \ge 0 \hspace{1mm}\forall \hspace{1mm}i$ i.e $r_1 \le r_2 \le r_3 \le ... \le r_n$. For that we are transforming the unconstrained vector ($u_1,u_2,...,u_n$) to constrained vector ($r_1,r_2,...,r_n$) by using nested logistic transformation -
 \;
 
\begin{center}
$\displaystyle r_k=T_k(u)=\frac{1}{1+\sum_{i=k}^{n}e^{-u_{i}}} \hspace{1mm}\forall \hspace{1mm} k=1,2,...,n$\;
\end{center}
 
\noindent Now by replacing ($r_1,r_2,...,r_n$) in our likelihood function and differentiating it w.r.t. each $u_i$, we get -
 \;
\begin{center}
 $\displaystyle\frac{\delta l}{\delta u_k} = -n_0\frac{\sum_{j=1}^{k}\frac{e^{-u_k}}{(1+\sum_{i=j}^{n}e^{-u_i})^2}}{1-\sum_{j=1}^{n}\frac{1}{1+\sum_{i=j}^{n}e^{-u_i}}}+\sum_{j=1}^{k}\frac{n_j.e^{-u_k}}{1+\sum_{i=j}^{n}e^{-u_i}}=0 \hspace{1mm}\forall\hspace{1mm}k=1,2,...,n$\;
\end{center}
 
\noindent To solve these equations we tried Sequential Quadratic Programming method for which we need to find the gradient and hessian matrix by differentiating the system of equations w.r.t $u_i$'s once and twice respectively. But it was very difficult to calculate. So we have tried another method.\;
 
 \subsection*{Fourth approach}
 We have seen that obtaining the probabilities analytically was difficult and problematic. So we shall now try numerical method. In this method also we shall try to use constrained MLE by calculating the $p_i$'s in iterative procedure. The steps are given below:\;
 
 \textbf{Step 1:} Define the likelihood function as per the obfuscation procedure.\;
 
 \textbf{Step 2:} Initialize $p_i$'s such that $p_i \ge 0 \hspace{1mm}\forall \hspace{1mm} i=1,2,...,n$ and $\sum p_i =1$\;
 
 \textbf{Step 3:} To determine the values of $p_i$'s we want the precision of the values of $p_i$'s upto third digit (say). In each step of the iteration we fix the sum and so we update $p_i$ by $sum \times \frac{1}{1000}$ and $p_{i+1}$ by $sum \times \frac{999}{1000}$ and check whether the value of the likelihood function is greater or not. In this way we shall update $p_i$ and $p_{i+1}$ by $sum \times \frac{j}{1000}$ and $p_{i+1}$ by $sum \times\frac{(1000-j)}{1000} \hspace{1mm} \forall \hspace{1mm} j=0,1,2,...,1000$ and compare with the likelihood value. In this way we shall get the values of $p_i$ and $p_{i+1}$ for which likelihood function will be maximum. We do the same thing for different values of i and repeat the whole process k times. Where k is sufficiently large.\;
 
 \;
 
\noindent\textbf{Correctness of the above process:} In the process we are keeping the sum fixed and vary the `j' to see for which proportion of `sum' should be $p_i$. So the sum of the $p_i$'s are always be 1 and they all are positive. [Thus we avoided the problems occurred in the previous methods.] In each step we are comparing the value of likelihood function in  current step with that of the previous step. And as we are taking k to be sufficiently large (i.e we are repeating the iterative process sufficient no. of times), the process will return the correct constrained MLE values.\;
 
 \vspace{5mm}
 
 \subsection*{Implementation} \;
 
 1. We first implement this procedure to the generated data set which we have used in previous approaches.\;
 
Here the log likelihood function is $L=n[1]\times log(x[1])+n[2]\times log(x[1]+x[2])+\cdots +n[11]\times log(x[1]+x[2]+x[3]+\cdots+x[11])+n[12]\times log(x[2]+x[3]+\cdots+x[12])+\cdots+ n[23]\times log(x[13])$\\
 
where, $x[i]$ be the probaility of $i^{th}$ class and $n[i]$ be the observed count in that class.\;

\vspace{3mm}
 
 [Necessary R codes are given in Appendix section.]\;

\;

 The results are as following -
 
 \begin{center}
     \includegraphics[scale=0.9]{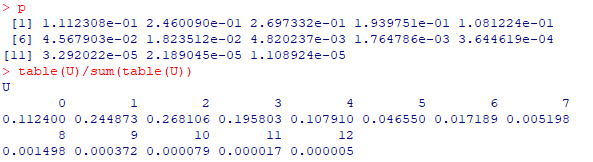}
 \end{center}
 
 The vector `p' gives the estimate of $p_i$'s using the iterative method and `table(u)/sum(table(u))' shows the proportions of values from actual data. We can observe that the estimation is quite well i.e the values in the two vectors are close enough. That's why quantiles can be estimated with high accuracy by taking the cumulative sums of the vector p.\\
 
 2. To check the consistency we generated a data set in the same way as the previous and implemented the procedure for that data set.\;
 
 The results are shown below:
 
 \begin{center}
     \includegraphics[scale=0.9]{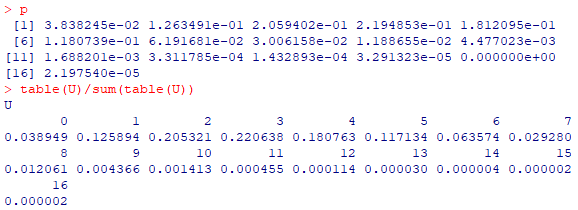}
 \end{center}
 
 Here also the vector `p' gives the estimate of $p_i$'s and `table(u)/sum(table(u))' shows the proportions of values from actual data. We can observe that in first few components the estimation is quite well but in last part the difference between the estimated values and the actual values are not that small and also there is a 0(zero) in $15^{th}$ component. To describe this observation we go through the diagram of obfuscated data. \\
 
 \begin{center}
     \includegraphics[scale=0.6]{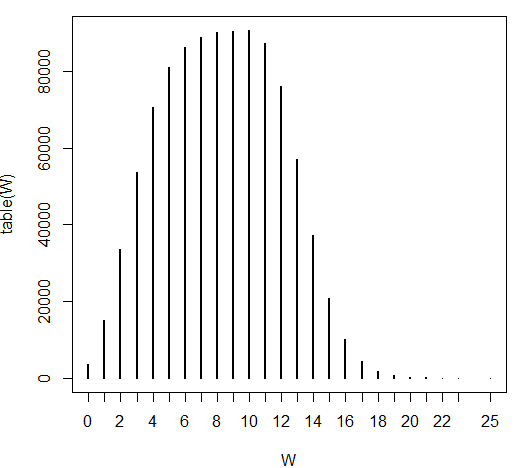}
 \end{center}
 
 We can notice that there is no observation for the value 24 which is affecting the likelihood function and consequently giving a poor estimate near to corresponding value in actual data.\;
 
 So for such cases we suggest to merge the class with any of the neighbour class and modify the likelihood function accordingly.\\
 
 3. Now we shall try to implement the procedure for the household size data set.\;
 
 Before implementing, we need to construct the raw data set from the summarize data. The main issue for construction is that there are some classes in the data set that contain a class of values and also the last class is unbounded. So we need to think about some techniques about breaking the classes in discrete points in order to construct the raw data set that we need. \;
 
 We have considered poisson distribution with parameter average household size (i.e 4.8) as the best fit for the data. We break those classes into discrete values according to the proportion of expected frequencies corresponding to each value.\;
 
 \vspace{5mm}
 
 After breaking the data set is like following:\;
 
  \begin{center}
     \includegraphics[scale=0.9]{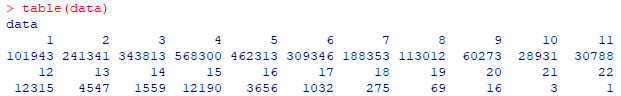}
 \end{center}
 
Here, the likelihood function is -\;

$L=n[1]\times log(x[1])+n[2]\times log(x[1]+x[2])+\cdots+n[10]\times log(x[1]+x[2]+\cdots+x[10])+n[11]\times log(x[2]+x[3]+\cdots+x[11])+\cdots+n[22]\times log(x[13]+x[14]+\cdots+x[22])+\cdots+n[29]\times log(x[20]+x[21]+x[22])$\\

where, $x[i]$ be the probaility of $i^{th}$ class and $n[i]$ be the observed count in that class.\\

The results are as following:\;

 \begin{center}
     \includegraphics[height=40 mm, width=156mm]{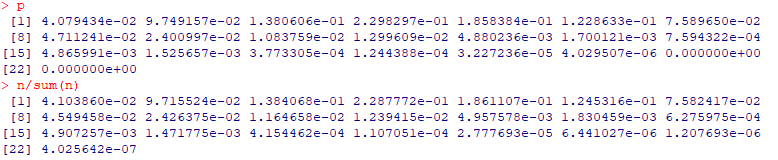}
 \end{center}

Similarly as the generated data for this household size data set also this procedure gives a quite good estimate of the distribution as we can see in the result.\;

\;

Here we are using the obfuscated data set for our estimation but to maintain the data privacy at extreme values we need to do truncation. So our next aim is to implement our procedure for the truncated data set.\\

We shall use our household size data set and will truncate the obfuscated data at 22. The truncated obfuscated data set will be as following:\\

\begin{center}
    \includegraphics[scale=0.6]{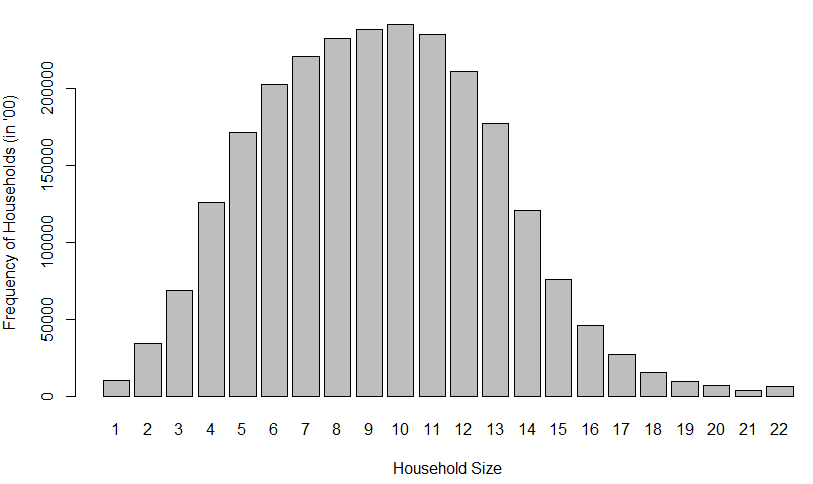}
\end{center}

In this case the log likelihood function is -$L=n[1]\times log(x[1])+n[2]\times log(x[1]+x[2])+\cdots+n[10]\times log(x[1]+x[2]+\cdots+x[10])+n[11]\times log(x[2]+x[3]+\cdots+x[11])+\cdots+n[21]\times log(x[12]+x[14]+\cdots+x[21])+n[22]\times log(10-10\times(x[1]+\cdots+x[12])- 9\times x[13] - \cdots - 2\times x[20] - x[21])$\\

where, $x[i]$ be the probaility of $i^{th}$ class and $n[i]$ be the observed count in that class.\\

The result are shown below -

\begin{center}
    \includegraphics[scale=0.95]{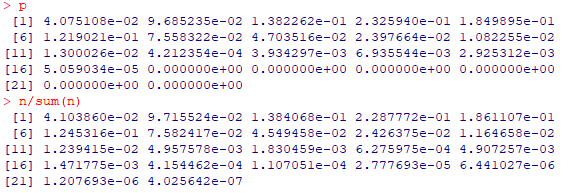}
\end{center}

As we can see from the result, the first few estimated probabilities are very close to that of actual data but the estimation gets worse at the end. This is because the fact that we truncated the observations at 22 and hence considering all the observations that have value more than or equal 22 as same quantity.\;

Although if we focus on estimation of quantiles, we proceed to check the cumulative sum of each vector and the result are as following -\;

\begin{center}
    \includegraphics[scale=0.95]{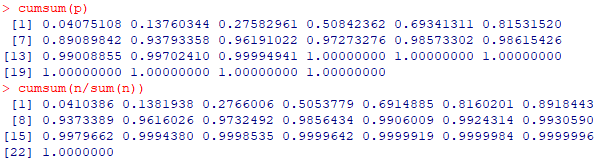}
\end{center}

As we can observe upto 99$^{th}$ percentile, the estimation is quite accurate. Only the last quantile can't be predicted due to the same reason of truncation.\;

Therefore, our procedure works fine if the data need to be published after truncation.\\

Point to be noted here, in the whole procedure we described, needed the likelihood function to be perfectly defined. Now to define the likelihood function, it is necessary to have a bit of knowledge of the range and if the range is published again we have compromise some data privacy. To maintain the privacy, we suggest to publish a wider range and we have checked that our procedure can be implemented with a knowledge of wider range also. \;

\;

\textbf{Implementation:} Here we consider the without truncation case where the actual range was (1,22), instead here we are publishing a wider range (1,25).\;

Therefore, for this case, log likelihood function is -

$L=n[1]\times log(x[1])+n[2]\times log(x[1]+x[2])+\cdots+n[10]\times log(x[1]+x[2]+\cdots+x[10])+n[11]\times log(x[2]+x[3]+\cdots+x[11])+\cdots+n[22]\times log(x[13]+x[14]+\cdots+x[22])+\cdots+n[29]\times log(x[20]+x[21]+x[22]+\cdots+x[25])$\\

where, $x[i]$ be the probaility of $i^{th}$ class and $n[i]$ be the observed count in that class.

\begin{center}
    \includegraphics[scale=0.8]{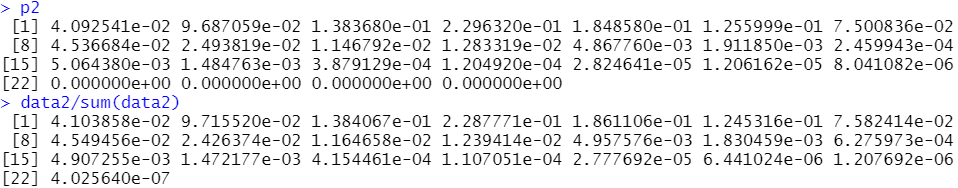}
\end{center}

So we can see from the result that even if we declare a wider range, the estimated probability are close to the actual value and the probabilities corresponding to values 23-25 which are not in the actual data coming to be zero as expected.\;

\section*{Goodness of estimates}

To test for the goodness of the estimates we compute the individual variance and MSEs of each component of the estimate. Bootstrapping from the predicted data we numerically generate multiple copies of the raw obfuscated data and numerically calculate the variances and MSEs. The computed vectors are as seen below:

\includegraphics[width=15cm,height=32mm]{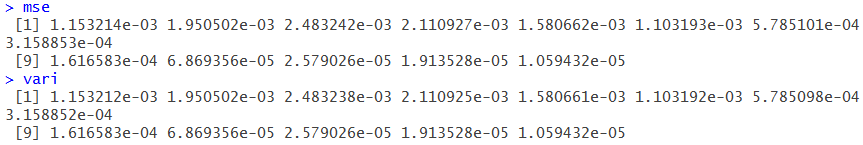}\;

We note that the variances are of the order $\mathcal{O}(10^{-3})$ and the corresponding MSEs are very close to them. Hence, we can say that the estimates are of low bias and variance and so it is consistent.\;

\section*{Approaches for estimating Range}

Since our actual goal is to hide the information about the raw data. So we want to see whether the estimate of range of the actual data can be estimated sufficiently close to the real value.\;\\

\noindent\textbf{Using Law of large number\footnote[5]{Nandi Mridul $\&$ Roy Bimal [2021] Estimation of extremum of obfuscated data and its error of estimation}:}\;

We first try to estimate the range from the data set that we generated from poisson distribution and added an additive noise Discrete Uniform (0,10) to it. Suppose`W’ is the obfuscated data. We shall add the same noise 999 times to the obfuscated data. So we have `W1’(say) which is a data after obfuscation the actual data 1000 times. If we consider 1000 as a large number, we know that the maximum element of actual data should be close to (the maximum element of `W1’-5*1000)[Using Law of large number]. But we know that the maximum should be 12 according to our data and we observe the results to be in between 450-500. The results in R are given below: \;
\begin{center}
    \includegraphics[scale=0.6]{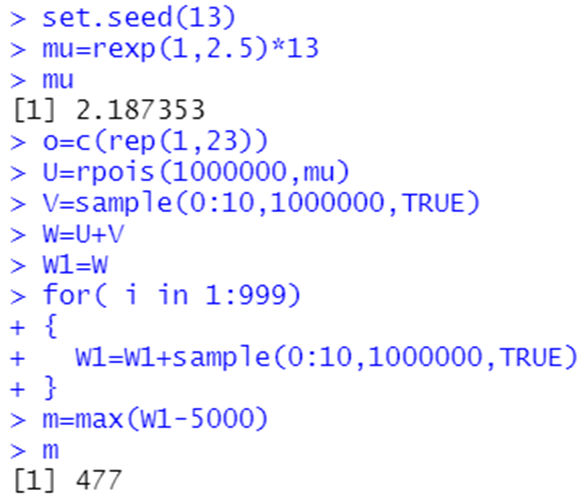}
\end{center}

We have also observed that if we increase the number of noise, the resulted maximum value is more far from that we expect.\;

\;

\subsection*{Another approach}

We know the minimum household size is 1. So here we are trying to estimate the maximum of the household size from the obfuscated data.\;

\textbf{Method of finding maximum:}\;

\textbf{Step 1:} Estimate the quantiles from obfuscated data by previously mentioned method (Fourth approach to estimate quantiles).\;

\textbf{Step 2:} The maximum is estimated as the value at $100^{th}$ percentile. It may happen that the $100^{th}$ percentile is not unique. In that case, the maximum will be estimated from the minimum value that is at $100^{th}$ percentile.\;

\;

\textbf{Implementation:} \;

\textbf{Case 1:} (Without Truncation)
\begin{center}
    \includegraphics[scale=0.9]{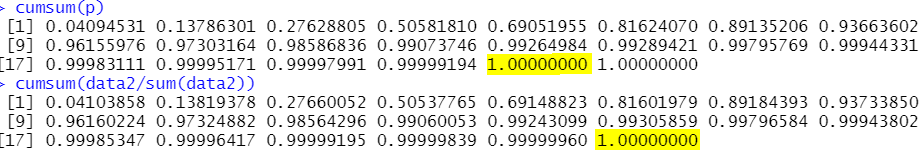}
\end{center}

By the procedure described above we check the $100^{th}$ percentile using our estimated probabilities and we can see that $100^{th}$ percentile occurs first at value 21. [The maximum value of the actual data is 22. So there is a small error in estimation.]\;

\;

\textbf{Case 2:} (Truncated)

\begin{center}
    \includegraphics[scale=0.9]{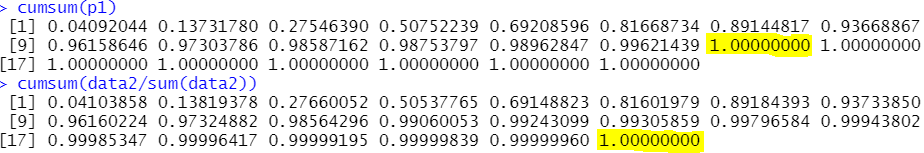}
\end{center}

We can see that in this case $100^{th}$ percentile occurs first at value 15, which is far from that of the actual value of maximum. This result was expected because we have truncated the obfuscated data at 22, that's why the values close to the end are given less priority. So, to estimate the maximum value from the truncated obfuscated data is not as easy as the previous case.\;

\section*{Conclusion}
\textbf{1.} The method that we have used to obfuscate our data i.e. the additive white noise model is a good way of masking a discrete data set i.e. probability of estimating the actual values of the data (e.g. household size) from the obfuscated data is very low.\\

\noindent\textbf{2.} Given real life data set related to household size, we have shown the actual probability with estimated probability for different household size in the following plot:
\begin{center}
    \includegraphics[scale=0.7]{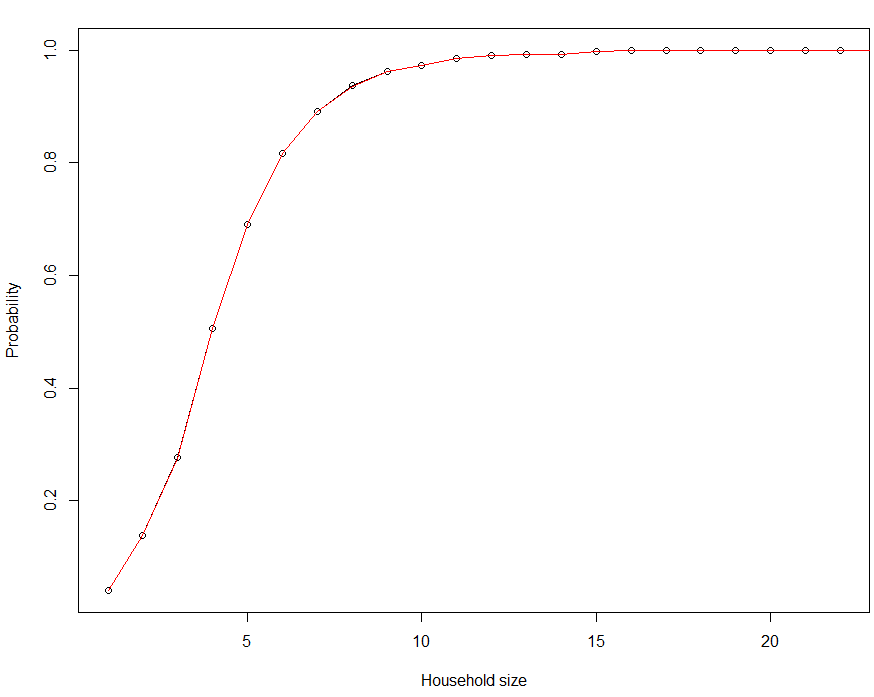}
\end{center}
We can notice in the above plot that the actual probability and the estimated values of the probabilities are very close to each other (the plots almost coincide with each other.\\

\noindent\textbf{3.} To estimate the probabilities from obfuscated data we tried Least square, MLE and constrained MLE methods. But there were some problems (e.g. estimate outside parameter space) which were solved in the fourth approach (numerical method). Also we have seen that the result obtained using iteration in this approach was very good.\\

\noindent\textbf{4.} While estimating the range of the raw data set, we first tried the method using LLN. But in that case the estimated values were very far from the actual value. Then we have tried another approach to estimate maximum value of the raw data set. In this method, we see that upto the $99^{th}$ percentile can be estimated quite accurately, however, the estimate for the $100^{th}$ percentile varies quite a lot from its real value. So there is further scope of study in this regard.\;

\pagebreak

\section*{Appendix:}

R codes for First Approach :
\begin{center}
\includegraphics[scale=0.6]{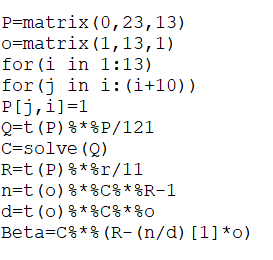}
\end{center}
R codes for the second approach :
\begin{center}
\includegraphics[scale=0.6]{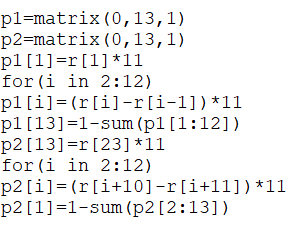}
\end{center}
\begin{center}
  \includegraphics[scale=0.6]{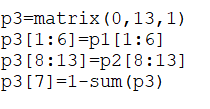}
\end{center}
R codes for Fourth approach :\;

Code for the generated data :
\begin{center}
    \includegraphics[scale=0.8]{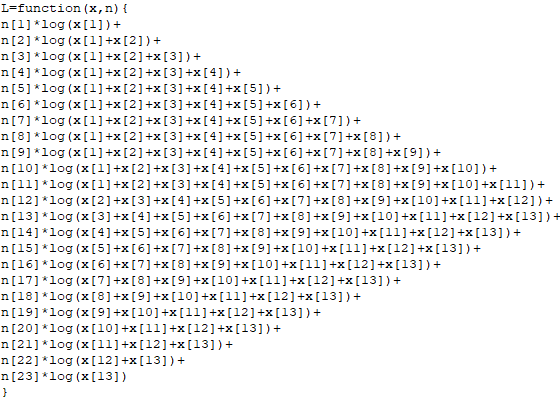}
    \includegraphics[scale=0.65]{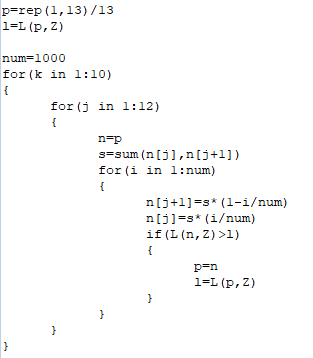}
\end{center}

Code for the non-truncated household size data :
\begin{center}       \includegraphics[scale=0.7]{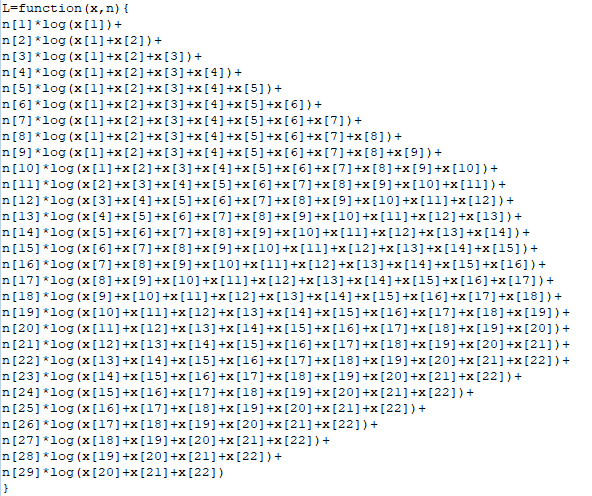}
\includegraphics[scale=0.85]{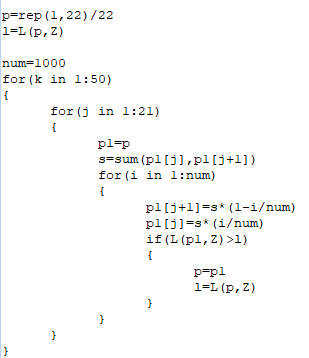}
 \end{center}
\pagebreak
Code for the truncated household size data :
\begin{center}
   \includegraphics[scale=0.8]{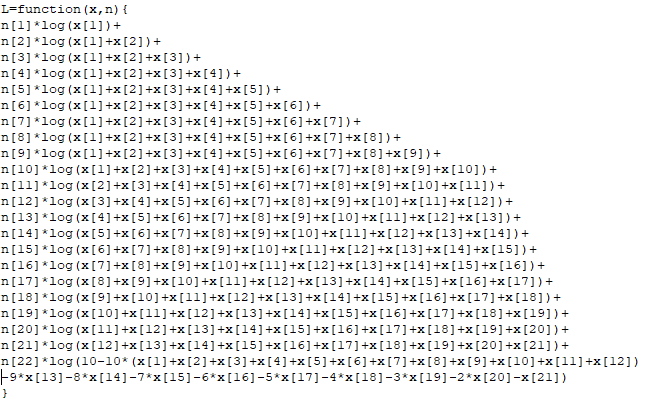}
   \includegraphics[scale=0.85]{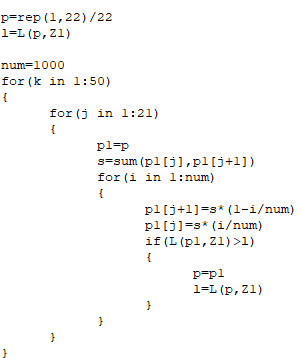}
\end{center}
\pagebreak
Code for the non-truncated household size data with wider range given:
\begin{center}
\includegraphics[scale=0.8]{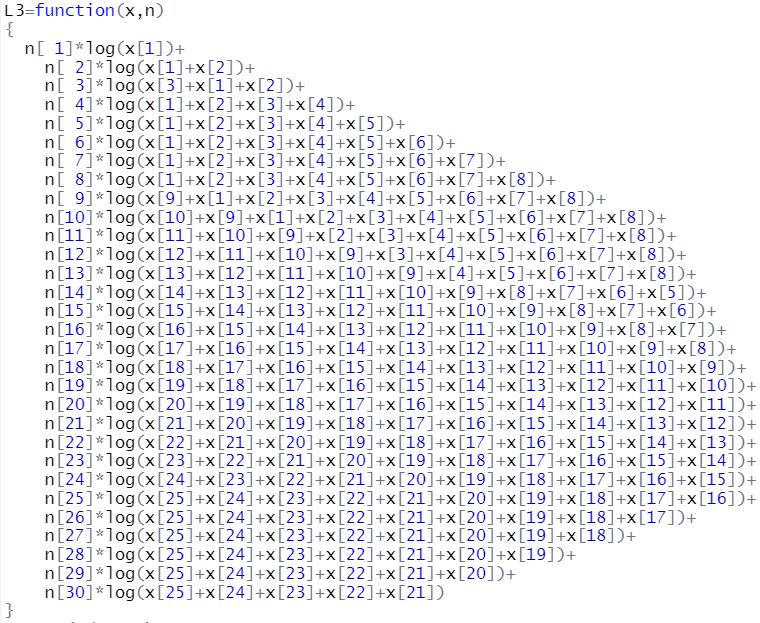}
\end{center}

\section*{Acknowledgements:}
We give our thanks to Prof. Bimal K. Roy, Prof. Nachiketa Chottopadhyay and Prof. Prabal Choudhury for their kind advice and support in this project. This project was also inspired by some previous works done by Prof. Bimal K. Roy et al. 

\section*{References :}
\begin{enumerate}
    \item Census of India, https://censusindia.gov.in/2011census/hh-series/hh01.html\;

    \item https://math.stackexchange.com/questions/2764174/how-to-solve-maximum-likelihood-estimates-with-inequality-constraint\;
    
    \item Ghatak, Debolina $\&$ Roy, Bimal. (2018). Estimation of True Quantiles from Quantitative Data Obfuscated with Additive Noise. Journal of Official Statistics. 34. 671-694. 10.2478/jos-2018-0032. 
    
    \item Nandi, Mridul $\&$ Roy, Bimal. (2021). Estimation of extremum of obfuscated data and its error of estimation.
\end{enumerate}

\end{document}